\documentclass{llncs}

\usepackage{graphicx}
\usepackage{booktabs}
\usepackage{multirow}
\usepackage{xspace}
\usepackage{url}
\usepackage{hyperref}
\usepackage{comment}
\usepackage{float}

\begin{document}

\title{Measuring Post-Quantum TLS Deployment Across UK Internet Sectors}

\titlerunning{Measuring Post-Quantum TLS Deployment Across UK Internet Sectors}
\author{Konstantinos Loizou \and Essam Ghadafi }

\authorrunning{K. Loizou and E. Ghadafi}

\institute{School of Computing, Newcastle University, Newcastle upon Tyne, UK\\
\email{k.loizou2@newcastle.ac.uk}\\
\email{essam.ghadafi@newcastle.ac.uk}}

\maketitle

\begin{abstract}
Post-quantum cryptography (PQC) is becoming an important component of
long-term trust in Internet-facing infrastructure. Publicly observable PQC
support provides evidence of externally visible deployment, but does not
necessarily reflect the overall progress of an organisation's post-quantum
migration. This distinction matters when observable deployment is used as an
indicator of organisational readiness or progress towards migration deadlines.

We present a measurement study of observable PQC deployment across 4,665 UK
organisations spanning ten sectors. We measure post-quantum key-exchange
support across HTTPS and SMTP STARTTLS endpoints, attribute reachable
endpoints to their underlying infrastructure providers where possible, and
statistically examine protocol-, sector-, and provider-level deployment
patterns. Among reachable endpoints, 44.0\% of HTTPS services supported at
least one evaluated PQC key-exchange group, compared with 6.4\% of SMTP
services. Among organisations reachable over both protocols, HTTPS support
was significantly more common than SMTP support (matched odds ratio 16.89).
Although deployment varied across sectors, infrastructure provider identity
was substantially more predictive than organisational sector, and observable
deployment was highly concentrated among a small number of providers. Only
144 organisations supported PQC across both web and email infrastructure,
highlighting an uneven and fragmented migration landscape. No post-quantum
certificate signatures were observed across the measured endpoints. These
findings show that observable PQC deployment is currently shaped
predominantly by infrastructure-provider deployment decisions and should not
be interpreted as a complete measure of organisational migration readiness.

\end{abstract}

\section{Introduction}
\label{sec:intro}
Public-key cryptography underpins the security of modern Internet services, supporting authentication, key establishment, and secure communications across web, email, and cloud infrastructures \cite{TLS13,RSA78,Mil85,Kob87}. However, the emergence of large-scale quantum computers threatens many of the asymmetric cryptographic algorithms currently deployed in practice. In particular, Shor's algorithm can efficiently solve the integer factorisation and discrete logarithm problems on which widely used schemes such as RSA and elliptic-curve cryptography rely \cite{Shor94}. The possibility of adversaries collecting encrypted traffic today for future decryption, commonly referred to as the \emph{harvest-now-decrypt-later} threat, further increases the urgency of migration towards quantum-resistant alternatives.

To address this challenge, significant international efforts have focused on standardising and deploying post-quantum cryptography (PQC). In 2024, the U.S. National Institute of Standards and Technology (NIST) standardised the first generation of post-quantum cryptographic algorithms, publishing ML-KEM, ML-DSA and SLH-DSA as FIPS 203, FIPS 204 and FIPS 205, respectively \cite{FIPS203,FIPS204,FIPS205}. In parallel, hybrid key exchange mechanisms have been proposed to facilitate the gradual integration of post-quantum cryptography into TLS while maintaining compatibility with existing deployments \cite{IETFHybridTLS}.

Governments and regulatory bodies have also begun establishing migration roadmaps for the transition to quantum-resistant cryptography. In the United Kingdom, the National Cyber Security Centre (NCSC) has published a phased migration timeline recommending that organisations complete cryptographic discovery, dependency mapping, and migration planning by 2028, followed by the progressive migration of systems and services during the subsequent years \cite{NCSCMigration}. Similarly, the European Commission has recommended a coordinated roadmap to support the transition to post-quantum cryptography across member states and critical sectors \cite{EUPQCStrategy24}. These initiatives highlight the growing importance of understanding the current state of PQC deployment and establishing baselines against which future migration progress can be assessed.

Recent industry developments further emphasise the urgency of understanding current deployment practices. Major Internet infrastructure providers have accelerated their own migration roadmaps, with Google and Cloudflare publicly targeting 2029 for completing the migration of their infrastructures to post-quantum cryptography \cite{google2026pqc,cloudflare2026roadmap}. Given the substantial proportion of Internet-facing services delivered through these providers, their deployment decisions are likely to have a significant influence on the observable state of PQC adoption across organisations. Consequently, understanding the contribution of infrastructure providers is important when interpreting Internet-scale measurements of organisational PQC readiness.

The United Kingdom provides a particularly relevant case study because the NCSC has published one of the most comprehensive national migration roadmaps for PQC adoption, allowing observable deployment to be interpreted within a well-defined policy framework.

Despite recent progress in measuring PQC deployment across Internet-facing services, several important challenges remain. Existing studies predominantly rely on popularity-ranked domain lists, provide limited visibility into sector-specific organisational deployment, typically analyse either HTTPS or SMTP in isolation, and provide limited insight into the influence of infrastructure providers on observable deployment.
To address these challenges, this paper presents an empirical measurement
study of 4,665 UK organisations spanning ten sectors. Using a sector-stratified dataset, we jointly analyse HTTPS and SMTP STARTTLS infrastructures, attribute reachable endpoints to their underlying infrastructure providers, and examine leaf certificate signature algorithms alongside post-quantum key exchange support. Collectively, these analyses provide a complementary perspective to existing Internet-wide measurement studies by focusing on organisational sectors and the role of infrastructure providers in shaping observable PQC deployment.

Accordingly, this study addresses the following research questions:
\begin{description}
    \item[RQ1:] How prevalent is observable PQC support across HTTPS and SMTP
    infrastructures?
    \item[RQ2:] Does observable PQC support vary across organisational sectors?
    \item[RQ3:] Is observable PQC support better predicted by organisational
    sector or infrastructure provider?
    \item[RQ4:] How concentrated is observable PQC deployment among
    infrastructure providers?
\end{description}

\noindent\textbf{Our Contribution.} The contributions of this paper are as follows:
\begin{itemize}
    \item We present a sector-stratified measurement study of observable PQC
deployment across 4,665 UK organisations spanning ten sectors.

    \item We jointly measure HTTPS and SMTP STARTTLS deployment for the same
    organisational domains, enabling paired comparison of observable PQC
    support across web and email infrastructures.

    \item We quantify differences in deployment across protocols and sectors
    using confidence intervals, paired hypothesis testing, and effect-size
    analysis.

    \item We compare the predictive value of organisational sector and
    infrastructure provider using cross-validated logistic regression,
    including MX-host-grouped validation for SMTP to account for shared mail
    infrastructure.

    \item We show that provider identity is substantially more predictive of
    observable PQC support than organisational sector, and that deployment is
    highly concentrated among a small number of infrastructure providers.

    \item As a secondary analysis, we characterise leaf certificate signature
    algorithms to assess whether visible deployment has progressed beyond
    post-quantum key establishment.
\end{itemize}

The remainder of the paper is organised as follows. Section~\ref{sec:relwork} reviews background and related work. Section~\ref{sec:method} describes the measurement methodology. Section~\ref{sec:results} presents the empirical results.
Section~\ref{sec:disc} discusses the implications of the findings for post-quantum migration and organisational readiness, and Section~\ref{sec:conclude} concludes the paper.

\section{Related Work}
\label{sec:relwork}
Large-scale Internet measurement has long been used to study the deployment and evolution of security protocols and public key infrastructures. Holz et al.~\cite{Holz2011SSL} presented one of the first comprehensive active and passive measurement studies of the X.509 Public Key Infrastructure (PKI), demonstrating the effectiveness of Internet-scale measurements for analysing certificate ecosystems and TLS deployments. More recently, Holz et al.~\cite{Holz2019TLS13} investigated the deployment of TLS~1.3 and showed that rapid protocol adoption was largely driven by a small number of cloud providers and content delivery networks. These observations highlight the importance of considering the role of infrastructure providers when interpreting Internet-wide deployment measurements. Other measurement studies have examined TLS deployment within specific sectors. For example, Chung and Vlajic analysed the TLS configurations of major banking websites using remote scanning techniques, highlighting the continued prevalence of legacy protocol versions and insecure cryptographic configurations in a critical sector \cite{Chung2022TLS}.

Several recent studies have investigated the deployment of post-quantum cryptography across Internet-facing services. Schaumann conducted longitudinal Internet measurement studies examining PQC deployment across Internet-facing websites and later extended the analysis to SMTP infrastructures \cite{SchaumannHTTPS2026,SchaumannSMTP2026}. Complementary evidence has also been reported in industry measurement reports, such as those published by F5 Labs, which analyse PQC deployment across large populations of Internet-facing websites and examine deployment trends across infrastructure providers and geographical regions \cite{F5PQC2025}. Collectively, these studies indicate increasing deployment of post-quantum TLS, although observable deployment remains highly concentrated among a relatively small number of major infrastructure providers.

While preparing this manuscript, a related measurement study of Internet PQC readiness became publicly available \cite{PQCReadiness2026}. That study analyses more than 32,000 Internet-facing domains by examining TLS protocol versions, cipher suites, key exchange mechanisms, and certificate configurations. Although complementary, the two studies differ in several important respects. Our work focuses specifically on UK organisations across 10 sectors, jointly analyses HTTPS and SMTP infrastructures, examines sector-level deployment patterns, attributes endpoints to their underlying infrastructure providers, and analyses certificate signature algorithms alongside post-quantum key exchange support. Rather than providing another Internet-wide scan, our study complements existing work by offering a sector-oriented perspective and investigating the influence of infrastructure providers on observable PQC deployment. Together, the two studies provide complementary perspectives on the current state of observable PQC deployment across Internet-facing services.
\section{Methodology}
\label{sec:method}
This section describes the methodology used to measure observable deployment
of post-quantum cryptography across UK organisations, including dataset
construction, HTTPS and SMTP measurements, infrastructure-provider
attribution, statistical analysis, and the ethical considerations that guided
the measurement campaign.

\subsection{Dataset Construction}
\label{sec:dataset}
The final measurement dataset comprises 4,665 UK organisations spanning ten
sectors. To provide broadly balanced representation while avoiding
over-representation of sectors containing substantially larger organisational
populations, the commercial-sector samples and the Government sample were
limited to a maximum of 500 organisations per sector. All 165 eligible Higher
Education institutions were included because this source contained fewer than
500 entries.

A sector-stratified sampling strategy was adopted using publicly available data
sources. Commercial organisations from the Construction, Utilities and
Environmental Services, Healthcare, Legal, Retail, Technology,
Telecommunications, and Finance and Insurance sectors were identified using
the Moody's FAME database \cite{MoodysFAME}. The dataset was filtered to
include only active organisations whose registered office was located in
England, Scotland, Wales, or Northern Ireland. Organisations were assigned to
sectors using their primary UK Standard Industrial Classification (SIC) code.
Table~\ref{tab:siccodes} summarises the primary SIC codes used to construct
the commercial-sector dataset.

\begin{table}[tbh]
\centering
\caption{Primary UK SIC codes used to construct the commercial-sector dataset.}
\label{tab:siccodes}
\small
\begin{tabular}{|l|l|}
\hline
\textbf{Sector} & \textbf{Primary UK SIC code(s)} \\
\hline
Construction & 41, 42, 43 \\
Finance \& Insurance & 64, 65, 66 \\
Healthcare & 86 \\
Legal & 691 \\
Retail & 47 \\
Technology & 62, 63 \\
Telecommunications & 61 \\
Utilities \& Environmental Services & 35, 36, 37, 38, 39 \\
\hline
\end{tabular}
\end{table}

For each commercial sector, the first 500 eligible records returned by the
corresponding FAME query were selected. The website recorded by FAME was
used as the organisation's public Internet domain. Records without a valid
public domain were skipped and replaced using the next eligible record returned
by the query. Government organisations were obtained from the official GOV.UK
domain list \cite{govuk_domains}, with the first 500 eligible non-duplicate
domains retained. Higher Education institutions were obtained from the UK
Learning Providers dataset \cite{UK_Learning_Providers}.

An initial Police category was constructed using the UK Police API
\cite{police_uk_api}, together with the British Transport Police domain.
However, because this category contained fewer than the minimum threshold of
50 organisations adopted for reporting sector-level statistics, it was
excluded from all analyses and public reporting.

Each organisation in the final dataset was represented by a single publicly
accessible Internet domain corresponding to its primary organisational website.
Duplicate domains across sectors were removed to ensure that each domain
appeared only once. Where a null, invalid, or previously included domain was
encountered, the next available eligible entry from the corresponding source was
used to maintain the target sample size. This process resulted in a final
analytical dataset of 4,665 organisations across ten sectors.

All HTTPS and SMTP measurements reported in this paper were collected during
a single measurement campaign on 30 June 2026. The dataset therefore
represents a cross-sectional snapshot of observable PQC deployment at that
point in time.

\subsection{HTTPS \& SMTP Measurements}

HTTPS services were evaluated using TLS handshakes on TCP port~443. Each endpoint was probed using TLS~1.3 handshakes offering the 4 post-quantum key exchange groups listed in Table~\ref{tab:pqcgroups}. Since PQC key exchange is negotiated through the TLS~1.3 \texttt{supported\_groups} extension, only endpoints successfully negotiating TLS~1.3 were evaluated for PQC capability. For each successful handshake, the negotiated TLS version, negotiated key exchange group, X.509 certificate issuer, and leaf certificate's signature algorithm were recorded for subsequent analysis.

SMTP infrastructure was evaluated independently using the STARTTLS extension on TCP port~25. For each organisation, the highest-priority Mail Exchange (MX) host was identified from its DNS MX records. As multiple organisations frequently share common mail infrastructure, duplicate MX hosts were removed prior to scanning, ensuring that each unique mail server was measured only once.
Each unique MX host was subsequently probed using the hybrid key exchange groups listed in Table~\ref{tab:pqcgroups}. The resulting measurements were then mapped back to all organisations sharing the corresponding mail infrastructure. Alongside PQC support, the negotiated TLS version, X.509 certificate issuer,
and leaf certificate signature algorithm were recorded for each successful
connection, enabling comparative analysis of HTTPS and SMTP deployment
across the dataset.

\begin{table}[thb]
\caption{TLS 1.3 post-quantum key exchange groups evaluated in this study.}
\label{tab:pqcgroups}
\centering
\begin{tabular}{|l|l|l|}
\hline
\multicolumn{1}{|c|}{\textbf{Key Exchange Group}} & \multicolumn{1}{|c|}{\textbf{Classical Component }} & \multicolumn{1}{|c|}{\textbf{PQC component}} \\
\hline
X25519MLKEM768        & X25519     & ML-KEM-768 \\
SecP256r1MLKEM768     & secp256r1  & ML-KEM-768 \\
MLKEM1024             & --         & ML-KEM-1024 \\
SecP384r1MLKEM1024    & secp384r1  & ML-KEM-1024 \\
\hline
\end{tabular}
\end{table}

The evaluated key exchange groups comprise the hybrid and pure post-quantum key exchange groups supported by the OpenSSL implementation used during data collection. The hybrid groups combine a classical elliptic-curve key exchange mechanism with ML-KEM, whereas \texttt{MLKEM1024} represents a pure post-quantum key exchange group. An endpoint was classified as supporting observable PQC deployment if at least
one of the evaluated post-quantum key-exchange groups was successfully
negotiated.

\subsection{Infrastructure Provider Attribution}
\label{sec:methodinfr}
Following completion of the HTTPS and SMTP measurements, a separate infrastructure attribution stage was performed to identify the underlying infrastructure provider responsible for each reachable endpoint.

For HTTPS services, provider attribution combined multiple complementary techniques, including DNS CNAME resolution, reverse DNS (PTR) lookups, Autonomous System (AS) mapping, WHOIS organisation-name matching and HTTP response header analysis. The implementation was based on the open-source \textit{whohosts} software~\cite{Schaumann_whohosts}, which integrates these attribution techniques into a unified workflow.

SMTP infrastructure was attributed by analysing the highest-priority MX hostname associated with each organisation. MX hostnames were matched against provider-specific patterns representing the major email service providers, including Microsoft, Google, Hostinger, Proofpoint, Mimecast, Barracuda, Cisco, Cloudflare, Apple iCloud, Amazon, Namecheap and other widely deployed hosted email services. Additional provider-specific patterns were incorporated for organisations operating multiple hosting brands, including United Internet, to improve attribution coverage. MX hostnames that could not be matched to a recognised provider were classified as \textit{Other}.

The resulting provider attributions were subsequently used to analyse the distribution of observable PQC deployment across major hosting and email infrastructure providers.


\subsection{Statistical Analysis}
\label{sec:stats}
Statistical analyses were performed to evaluate deployment patterns at the
protocol, sector, and infrastructure-provider levels. Deployment proportions
are reported with Wilson 95\% confidence intervals. Differences between HTTPS
and SMTP deployment were evaluated among organisations reachable over both
protocols using McNemar's test with continuity correction. The corresponding
matched odds ratio and 95\% confidence interval were calculated using the
discordant pairs.

Associations between organisational sector and observable PQC deployment were
evaluated using Pearson's chi-square test of independence, with Cramer's $V$
reported as a measure of effect size. These analyses used the final ten-sector
dataset after exclusion of the Police category described in
Section~\ref{sec:dataset}.

To compare the predictive value of organisational sector and infrastructure
provider, logistic regression models were fitted using three predictor sets:
(i) sector only, (ii) provider only, and (iii) sector and provider combined.
Categorical predictors were one-hot encoded, and model discrimination was
measured using the area under the receiver operating characteristic curve
(AUC). HTTPS models were evaluated using stratified five-fold
cross-validation. SMTP models were evaluated using five-fold grouped
cross-validation, with organisations grouped by anonymised MX-host identifier
so that organisations sharing the same mail infrastructure were not divided
between training and test folds.

Provider concentration was summarised by ranking infrastructure providers
according to the number of PQC-supporting endpoints attributed to each
provider. Concentration was then quantified using the proportions
attributable to the highest-ranked provider, the three highest-ranked
providers, and the Herfindahl--Hirschman Index (HHI).

All analyses were implemented in Python~3.13 using
\texttt{pandas}, \texttt{NumPy}, \texttt{SciPy}, \texttt{statsmodels}, and
\texttt{scikit-learn}.


\subsection{Ethical Considerations}
The study collected only publicly observable TLS handshake metadata from Internet-facing HTTPS and SMTP services. To minimise operational impact, the measurements employed conservative scanning parameters, including connection timeouts, inter-connection delays and rate-limiting mechanisms. Each endpoint received only the minimum number of standards-compliant TLS handshakes required to determine protocol reachability, supported post-quantum key exchange groups and certificate characteristics, and all connections were terminated cleanly following protocol completion.

The measurement methodology was designed in accordance with the ethical principles outlined in the Menlo Report \cite{MenloReport} by minimising potential disruption while collecting only the metadata necessary to assess observable deployment of post-quantum cryptography. No attempts were made to authenticate to services, exploit vulnerabilities,
enumerate additional resources, or otherwise interfere with the operation of
the measured systems. A minimum sector size of 50 organisations was applied
for public reporting, and the Police category was excluded because it did not
meet this threshold. Organisation domains and MX hostnames were replaced with stable pseudonymous
identifiers for statistical analysis. Only aggregate statistics are reported in
this paper. The original scan data containing organisation identifiers are not
included in the published results and are retained only in a secure,
access-controlled environment for verification purposes.
\section{Results}
\label{sec:results}
This section presents the results of our empirical measurement study of post-quantum cryptographic deployment across UK organisations.

\subsection{Measurement Coverage}
The final analytical dataset comprised 4,665 UK organisations spanning ten
sectors. During the measurement campaign, 4,063 organisations (87.10\%) were
reachable over HTTPS, while 3,858 organisations (82.70\%) were reachable over
SMTP.

The SMTP dataset contained 2,615 distinct resolvable MX-host identifiers.
Among the SMTP-reachable organisations used in the predictive analysis, the
3,858 observations corresponded to 2,408 distinct MX-host groups. Adoption
statistics are reported at the organisation level to enable comparison with
HTTPS and sector-level analysis, while the grouped cross-validation procedure
described in Section~\ref{sec:stats} accounts for organisations sharing the same
mail infrastructure.

\begin{table}[tbh]
\centering
\caption{Summary of measurement coverage.}
\label{tab:coverage}
\begin{tabular}{|l|c|c|}
\hline
\textbf{Metric} & \textbf{HTTPS} & \textbf{SMTP} \\
\hline
Organisations in analytical dataset & 4,665 & 4,665 \\
Reachable endpoints & 4,063 & 3,858 \\
Reachability (\%) & 87.10 & 82.70 \\
Distinct resolvable MX hosts & -- & 2,615 \\
\hline
\end{tabular}
\end{table}


\subsection{Observable PQC Deployment Across Protocols and Sectors}
Table~\ref{tab:https-pqc-groups-sector} presents the distribution of the
evaluated HTTPS post-quantum key-exchange groups across the ten measured
sectors. The \texttt{X25519MLKEM768} hybrid group was by far the most widely
deployed, with 1,787 HTTPS endpoints supporting this configuration.
Government exhibited the highest number of deployments (255), followed by
Retail (230), Technology (212), and Legal (188).

Support for the remaining groups was considerably less common. A total of 161
HTTPS endpoints supported \texttt{SecP256r1MLKEM768}, while the pure
post-quantum \texttt{MLKEM1024} group was observed at only three endpoints.
The \texttt{SecP384r1MLKEM1024} group was supported by 59 endpoints. Overall,
1,788 of the 4,063 reachable HTTPS endpoints supported at least one evaluated
post-quantum key-exchange group. Observable HTTPS deployment was therefore
overwhelmingly centred on \texttt{X25519MLKEM768}, while alternative hybrid
configurations and pure post-quantum key exchange remained rarely deployed.

SMTP exhibited substantially lower PQC deployment and algorithm diversity than
HTTPS. Across the ten sectors, 246 SMTP endpoints supported at least one
evaluated post-quantum key-exchange group. Support was almost entirely limited
to \texttt{X25519MLKEM768}; only one SMTP endpoint, in the University sector,
also supported \texttt{SecP256r1MLKEM768}. Technology exhibited the highest
number of PQC-supporting SMTP endpoints (72), followed by Retail (57) and
Telecommunications (30). Observable SMTP PQC deployment therefore remained
considerably less prevalent and less diverse than HTTPS deployment.

Table~\ref{tab:https-smtp-pqc-both} presents the number of organisations
supporting at least one evaluated post-quantum key-exchange group across both
HTTPS and SMTP. Although 1,788 HTTPS endpoints and 246 SMTP endpoints
supported PQC, only 144 organisations supported PQC across both services.
Technology (42 organisations) and Retail (40) exhibited the highest levels of
dual-service deployment, followed by Legal (14) and Telecommunications (12).
Deployment across multiple Internet-facing services therefore remained
comparatively limited.

\begin{figure}[tbh]
    \centering
    \includegraphics[width=0.98\textwidth]{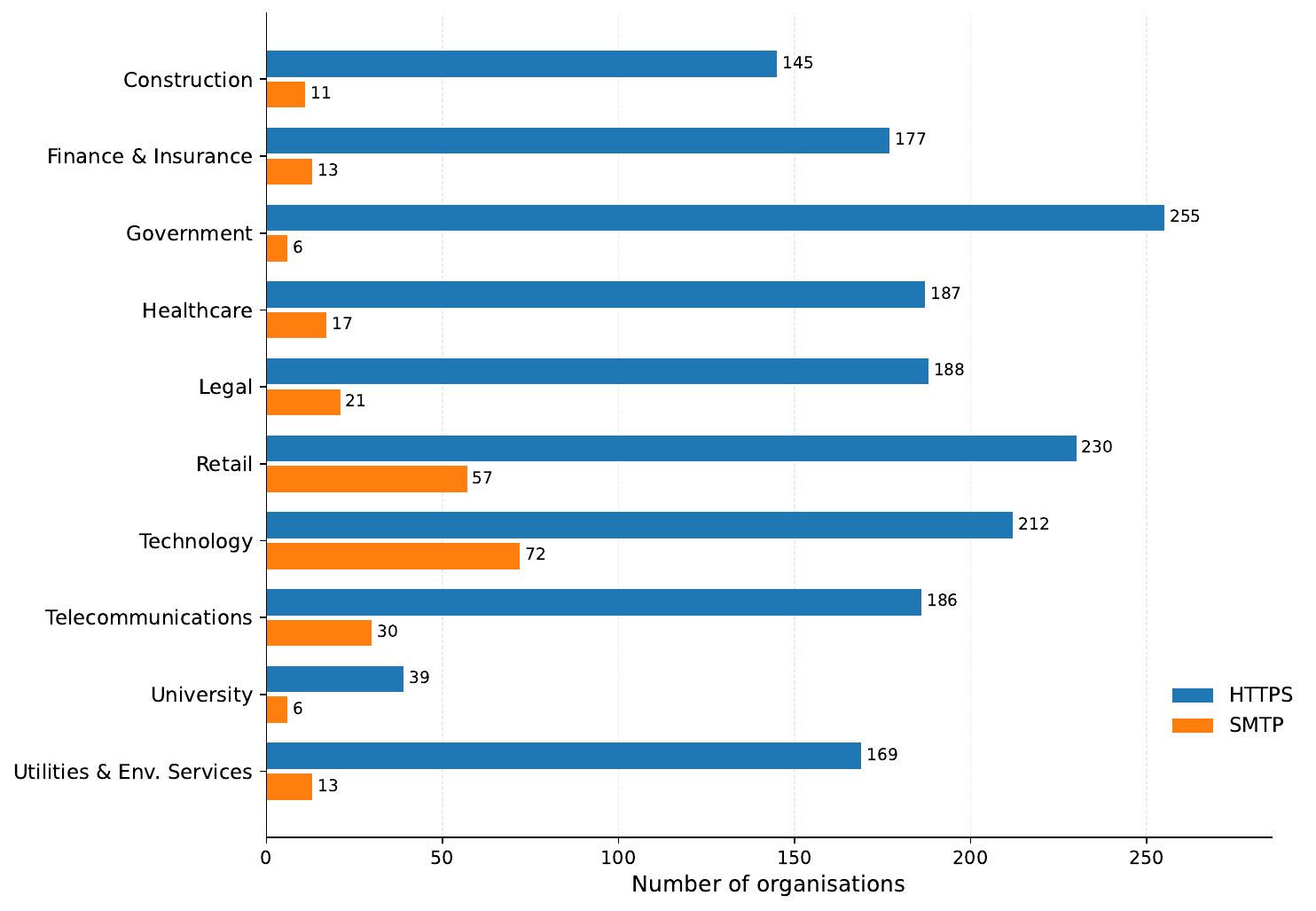}
    \caption{Number of organisations supporting  at least one of the evaluated post-quantum key exchange groups over HTTPS and SMTP across the 10 UK sectors included in the analytical dataset.}
    \label{fig:sector-pqc}
\end{figure}

Figure~\ref{fig:sector-pqc} compares observable PQC support over HTTPS and
SMTP across the ten sectors. HTTPS deployment consistently exceeded SMTP
deployment across all sectors, although the magnitude of this difference
varied considerably. SMTP deployment remained much more limited, with
Technology, Retail, and Telecommunications exhibiting the largest numbers of
PQC-supporting endpoints.

To address RQ1, we compared the prevalence of observable PQC deployment
across the two protocols. Among reachable endpoints, 1,788 of 4,063 HTTPS
services supported at least one evaluated post-quantum key-exchange group,
corresponding to 44.0\% (95\% Wilson CI: 42.5--45.5\%). For SMTP, 246 of
3,858 reachable services supported PQC, corresponding to 6.4\% (95\% Wilson
CI: 5.6--7.2\%).

Restricting the paired analysis to the 3,510 organisations reachable over both
protocols, 144 supported PQC over both services, 1,419 supported PQC only over
HTTPS, 84 supported PQC only over SMTP, and 1,863 supported neither.
McNemar's test confirmed that observable PQC support was significantly more
common over HTTPS than SMTP ($\chi^2=1184.00$, $p<0.001$). The matched
odds ratio was 16.89 (95\% CI: 13.56--21.05).

\begin{table}[t]
\caption{Support for the evaluated HTTPS post-quantum key exchange groups by sector. The total number of reachable HTTPS endpoints in each sector is shown in parentheses. ``Env. Serv.'' denotes Environmental Services.}
\label{tab:https-pqc-groups-sector}
\centering
\begin{tabular}{|l|c|c|c|c|}
\hline
\textbf{Sector} &
\shortstack{\textbf{X25519}\\\textbf{MLKEM768}} &
\shortstack{\textbf{SecP256r1}\\\textbf{MLKEM768}} &
\shortstack{\textbf{MLKEM}\\\textbf{1024}} &
\shortstack{\textbf{SecP384r1}\\\textbf{MLKEM1024}} \\
\hline
Construction (445) & 145 & 13 & 0 & 7 \\
Finance \& Insurance (425) & 177 & 25 & 0 & 1 \\
Government (455) & 255 & 1 & 0 & 1 \\
Healthcare (445) & 186 & 11 & 0 & 3 \\
Legal (430) & 188 & 13 & 1 & 12 \\
Retail (454) & 230 & 28 & 0 & 4 \\
Technology (437) & 212 & 31 & 1 & 6 \\
Telecommunications (407) & 186 & 10 & 1 & 1 \\
University (148) & 39 & 5 & 0 & 4 \\
Utilities \& Env. Serv. (417) & 169 & 24 & 0 & 20 \\
\hline
\textbf{Overall (4063)} & \textbf{1787} & \textbf{161} & \textbf{3} & \textbf{59} \\
\hline
\end{tabular}
\end{table}

To address RQ2, Pearson's chi-square tests indicated statistically
significant associations between organisational sector and observable PQC
deployment for both HTTPS ($\chi^2(9)=85.01$, $p<0.001$) and SMTP
($\chi^2(9)=165.10$, $p<0.001$). The corresponding effect sizes were modest,
with Cramer's $V=0.145$ for HTTPS and $V=0.207$ for SMTP. Although sector was
associated with observable deployment, the relatively small effect sizes
indicate that sector alone provides limited discrimination between
PQC-supporting and non-supporting endpoints.

Overall, these findings demonstrate that observable PQC deployment is substantially more prevalent for HTTPS than SMTP. Although observable deployment varied significantly across organisational sectors, the modest effect sizes suggest that organisational sector alone provides limited explanatory power, motivating the provider-level analysis presented in the following section.

\begin{table}[tb]
\caption{Number of organisations supporting at least one of the evaluated post-quantum key exchange groups across both HTTPS and SMTP services, grouped by sector.}
\label{tab:https-smtp-pqc-both}
\centering
\begin{tabular}{|l|c|}
\hline
\textbf{Sector} & \textbf{Domains} \\
\hline
Construction & 5 \\
Finance \& Insurance & 11 \\
Government & 2 \\
Healthcare & 7 \\
Legal & 14 \\
Retail & 40 \\
Technology & 42 \\
Telecommunications & 12 \\
University & 2 \\
Utilities \& Environmental Services & 9 \\
\hline
\textbf{Overall} & \textbf{144} \\
\hline
\end{tabular}
\end{table}

\subsection{Deployment Across Infrastructure Providers}
Tables~\ref{tab:https-providers} and~\ref{tab:smtp-providers} summarise the
major infrastructure providers ranked by the number of reachable endpoints
attributed to each provider.

\begin{table}[tbh]
\centering
\caption{Top 10 HTTPS infrastructure providers ranked by the number of reachable endpoints. Percentages indicate the proportion of reachable endpoints supporting at least one evaluated post-quantum key-exchange group.}
\label{tab:https-providers}
\small
\begin{tabular}{|l|c|c|c|}
\hline
\textbf{Provider} & \textbf{Reachable} & \textbf{PQC} & \textbf{PQC (\%)} \\
\hline
Cloudflare   & 1312 & 1247 & 95.05 \\
Amazon AWS   & 606  & 228  & 37.62 \\
Microsoft    & 261  & 4    & 1.53 \\
Google       & 176  & 12   & 6.82 \\
Akamai       & 143  & 2    & 1.40 \\
20i          & 138  & 0    & 0.00 \\
Krystal      & 121  & 0    & 0.00 \\
DigitalOcean & 116  & 2    & 1.72 \\
Fastly       & 81   & 78   & 96.30 \\
Ionos        & 72   & 1    & 1.39 \\
\hline
\end{tabular}
\end{table}

HTTPS deployment was concentrated among a small number of providers.
Cloudflare hosted 1,312 reachable HTTPS endpoints, of which 1,247 (95.05\%)
supported at least one evaluated post-quantum key-exchange group. Fastly also
exhibited a high deployment rate, with 78 of 81 endpoints (96.30\%) supporting
PQC. In contrast, several providers hosting substantial numbers of reachable
endpoints exhibited little or no observable deployment.

\begin{table}[tbh]
\caption{Top 10 SMTP infrastructure provider categories ranked by the number of reachable endpoints. Eleven provider categories are shown because of a tie for the tenth position. Percentages are calculated using only reachable SMTP endpoints attributed to each provider category.}
\label{tab:smtp-providers}
\centering
\begin{tabular}{|l|c|c|c|}
\hline
\textbf{Provider} & \textbf{Reachable} & \textbf{PQC} & \textbf{PQC (\%)}\\
\hline
Microsoft & 1714 & 0 & 0.00\\
Other Providers & 739 & 3 & 0.41\\
Mimecast & 650 & 0 & 0.00\\
Proofpoint & 368 & 0 & 0.00\\
Google & 242 & 242 & 100.00\\
Barracuda & 91 & 0 & 0.00\\
Cisco & 25 & 0 & 0.00\\
Cloudflare & 10 & 0 & 0.00\\
Amazon & 5 & 0 & 0.00\\
Namecheap & 4 & 0 & 0.00\\
Zoho Mail & 4 & 0 & 0.00\\
\hline
\end{tabular}
\end{table}

The SMTP ecosystem exhibited a different profile. Microsoft hosted 1,714
reachable SMTP endpoints, none of which supported the evaluated PQC groups.
By contrast, all 242 Google-hosted SMTP endpoints supported PQC and accounted
for almost all observable SMTP deployment.

\begin{table}[tbh]
\centering
\caption{Cross-validated AUCs of logistic regression models predicting
observable PQC support.}
\label{tab:model-performance}
\begin{tabular}{|l|c|c|c|}
\hline
\textbf{Protocol} & \textbf{Sector only} & \textbf{Provider only} &
\textbf{Sector + Provider} \\
\hline
HTTPS & 0.573 & 0.957 & 0.962 \\
SMTP  & 0.156 & 0.994 & 0.996 \\
\hline
\end{tabular}
\end{table}

To address RQ3, Table~\ref{tab:model-performance} compares logistic regression models using organisational sector only, infrastructure provider only, and both predictors. For HTTPS, the
provider-only model achieved an AUC of 0.957, substantially outperforming the
sector-only model (AUC = 0.573). Combining sector and provider produced only
a marginal improvement (AUC = 0.962).

For SMTP, five-fold cross-validation was grouped by anonymised MX-host
identifier. The provider-only model achieved an AUC of 0.994, whereas the
sector-only model achieved an AUC of 0.156. Combining sector and provider
produced an AUC of 0.996. For comparison, ordinary row-wise stratified
cross-validation produced an SMTP sector-only AUC of 0.699; this result does
not account for organisations sharing the same mail infrastructure and is
therefore not used as the primary estimate.

To address RQ4, we quantified the concentration of observable deployment among
providers. Cloudflare accounted for 69.7\% of all PQC-supporting HTTPS
endpoints, while Cloudflare, Amazon AWS, and Fastly together accounted for
86.9\%. SMTP deployment was even more concentrated: Google accounted for
98.4\% of PQC-supporting SMTP endpoints, and the three largest provider
categories accounted for all observed SMTP deployment.

The concentration of infrastructure providers across all reachable endpoints
was also reflected by the Herfindahl--Hirschman Index (HHI), which was 1,378
for HTTPS and 2,797 for SMTP. These values indicate a substantially more
concentrated provider landscape for SMTP than for HTTPS.

These findings show that infrastructure-provider identity provides
substantially greater predictive power than organisational sector and that
observable PQC deployment is currently concentrated among a small number of
providers.

\subsection{Leaf Certificate Signing Algorithm }
\label{sec:cert}
In addition to evaluating post-quantum key exchange support, we analysed the signature algorithm used by the leaf certificate presented during successful TLS handshakes. This enables us to assess whether observable PQC deployment has progressed beyond key establishment to certificate-based authentication.

\begin{table}[t]
\centering
\caption{Leaf certificate signature algorithms observed across reachable
HTTPS and SMTP endpoints.}
\label{tab:certificate-signatures}
\begin{tabular}{|l|c|c|}
\hline
\textbf{Signature Algorithm} & \textbf{HTTPS} & \textbf{SMTP} \\
\hline
SHA-256 with RSA & 2561 & 3824 \\
ECDSA with SHA-256 & 940 & 0 \\
ECDSA with SHA-384 & 540 & 24 \\
Other RSA & 22 & 10 \\
\hline
Total certificates & 4063 & 3858 \\
\hline
\end{tabular}
\end{table}

Table~\ref{tab:certificate-signatures} summarises the leaf certificate signature algorithms observed in the certificates presented by all reachable HTTPS and SMTP endpoints.

RSA-based signatures dominate both protocols, with SHA-256 with RSA accounting for the majority of certificates. ECDSA is also widely deployed for HTTPS but are comparatively uncommon for SMTP. No post-quantum certificate signature algorithms were observed, indicating that, although hybrid post-quantum key exchange has begun to emerge, certificate-based authentication remains entirely dependent on classical cryptographic signature algorithms.

\section{Discussion}
\label{sec:disc}
The results presented in this study provide an empirical view of the current state of observable PQC deployment across UK Internet-facing organisations. Although PQC support is becoming increasingly visible, particularly for HTTPS infrastructures, deployment remains highly uneven across services, sectors and infrastructure providers. Consequently, observable protocol support should be interpreted as evidence of externally visible infrastructure capability rather than a comprehensive measure of organisational migration readiness.

\subsection{Interpretation of the Findings}
The results demonstrate a substantial disparity between observable PQC
deployment over HTTPS and SMTP. Of the reachable endpoints, 1,788 supported
at least one evaluated PQC group over HTTPS, compared with 246 over SMTP,
and only 144 organisations supported PQC across both services. The paired
analysis confirmed that observable support was approximately 16.9 times more
likely over HTTPS than SMTP. Evaluating a single protocol in isolation may therefore overstate an organisation's broader level of externally visible post-quantum readiness. Observable deployment should therefore be interpreted as evidence of externally visible capability rather than evidence that an organisation has completed PQC migration.

Observable deployment was overwhelmingly centred on
\texttt{X25519MLKEM768}. The remaining evaluated hybrid groups exhibited
limited deployment, while the pure post-quantum \texttt{MLKEM1024} group was
observed at only three HTTPS endpoints and was not observed over SMTP. Current
deployment therefore remains focused on hybrid migration mechanisms that
preserve a classical component.

Sector was significantly associated with observable deployment for both
protocols, but the corresponding effect sizes were modest. Cross-validated
modelling provided a stronger distinction: infrastructure-provider identity
substantially outperformed organisational sector as a predictor, and adding
sector to a provider-based model produced only marginal improvement. Raw
sector differences should therefore not be interpreted primarily as evidence of
sector-specific migration programmes; they appear to reflect, to a substantial
extent, the provider ecosystems used by organisations in those sectors.

Observable deployment was also highly concentrated. Cloudflare accounted for
69.7\% of PQC-supporting HTTPS endpoints, while Google accounted for 98.4\%
of PQC-supporting SMTP endpoints. Several other providers hosting substantial
numbers of reachable services exhibited little or no observable support. These findings indicate that present Internet-facing deployment is strongly shaped by provider-level configuration decisions rather than uniform organisation-specific adoption of post-quantum cryptography.

\subsection{Comparison with Previous Work}
The findings broadly agree with previous Internet-scale measurement studies reporting increasing deployment of post-quantum TLS across Internet-facing services. Schaumann's longitudinal measurements similarly observed that deployment is concentrated among organisations using major cloud and content delivery providers, while industry reports published by F5 Labs also identified Cloudflare as one of the principal drivers of observable HTTPS PQC deployment. Our results reinforce these observations by demonstrating that deployment remains highly concentrated among a relatively small number of infrastructure providers, particularly for HTTPS, while also showing that observable SMTP deployment remains considerably more limited.

During the preparation of this manuscript, Dubey and Varshney \cite{PQCReadiness2026}
published a large-scale measurement study examining PQC readiness across more than
32,000 Internet-facing domains. While both studies investigate observable deployment
of post-quantum cryptography, the two works differ substantially in scope and
methodology and should be considered complementary. Their dataset is drawn primarily
from the Tranco global domain popularity ranking, supplemented with a set of Indian
banking-sector domains, rather than using a nationally or sector-stratified sampling
methodology. In contrast, our work employs sector-stratified sampling using UK
Standard Industrial Classification (SIC) codes across the full analytical dataset,
jointly analyses HTTPS and SMTP infrastructures, attributes endpoints to their
underlying infrastructure providers, and evaluates deployment patterns using formal
statistical testing and cross-validated predictive modelling. A further methodological
distinction is that their measurements primarily analyse negotiated TLS parameters
obtained using browser-based clients, reflecting the interaction between client
capabilities and server configuration, whereas our measurements probe server-side
protocol support directly. These represent complementary perspectives on observable
PQC deployment, and together the two studies provide a broader picture of the current
state of Internet-wide post-quantum readiness.

\subsection{Implications for Organisational Readiness}
The findings have several implications for assessing progress towards national and international post-quantum migration roadmaps. The migration guidance published by the UK National Cyber Security Centre (NCSC) and the European Commission emphasises activities such as cryptographic discovery, dependency identification, migration planning and risk assessment before large-scale deployment of post-quantum cryptography. Observable protocol support alone cannot demonstrate that these activities have been completed, nor can it provide evidence that an organisation has comprehensively migrated its cryptographic infrastructure.

At the same time, the results demonstrate the important role played by infrastructure providers in accelerating Internet-wide deployment. Organisations relying on providers that have enabled hybrid post-quantum key exchange may benefit from improved protection of their externally visible services without necessarily modifying their own public-facing applications or infrastructure. Consequently, infrastructure providers are likely to play a significant role in the early phases of post-quantum migration by enabling widespread deployment across large numbers of organisations.

The certificate analysis provides an additional perspective on migration readiness. Although hybrid post-quantum key exchange was observed across a substantial number of HTTPS deployments, no post-quantum certificate signature algorithms were identified within the measured dataset. This suggests that current observable deployments remain focused on protecting key establishment, while certificate-based authentication continues to rely entirely on classical cryptographic algorithms. Observable deployment should therefore be interpreted as an indicator of externally visible infrastructure capability rather than as evidence of a complete transition to post-quantum cryptography.

Future Internet-scale measurement studies should continue to combine protocol measurements with infrastructure-provider attribution and additional deployment characteristics to provide a more comprehensive assessment of post-quantum migration progress and to distinguish provider-driven deployment from organisation-driven migration.

\subsection{Limitations}
Several limitations should be considered when interpreting the results. First,
the study focuses exclusively on UK organisations and therefore may not reflect
deployment characteristics in other geographical regions. Second, the
measurements consider only publicly observable HTTPS and SMTP services,
excluding internally deployed systems and other communication protocols.
Third, although the provider-attribution methodology successfully identifies
the major hosting and email infrastructure providers responsible for much of
the observed deployment, it cannot guarantee complete attribution for every
endpoint. Consequently, the provider statistics presented in this paper should
be interpreted as an approximation of infrastructure concentration rather than
a definitive attribution of organisational responsibility.

Fourth, the study evaluates only the post-quantum key-exchange groups
supported by the measurement platform and OpenSSL implementation used during
data collection. As post-quantum standards and implementations continue to
evolve, future studies should consider additional algorithms and deployment
configurations as they become supported by mainstream TLS implementations.

The commercial-sector samples are not probability samples of all UK
organisations. For each FAME query, the first 500 eligible records returned by
the database were retained, with null or duplicate domains replaced by
subsequent eligible records. The results should therefore be interpreted as
measurements of the constructed sector-stratified dataset rather than unbiased
population estimates for every UK organisation in each sector.

A minimum public-reporting threshold of 50 organisations was applied,
resulting in the exclusion of the Police sector from the reported analyses.
The paper therefore makes no claims about observable PQC deployment within
that sector. In addition, all measurements were collected on 30 June 2026 and
represent a single cross-sectional snapshot. The results cannot establish
deployment trends, growth rates, or persistence over time.

The sector-level SMTP analysis treats organisations as independent
observations. Because multiple organisations may share common mail
infrastructure, the reported significance levels should be interpreted with
appropriate caution. However, the corresponding effect size remained modest,
and the subsequent provider-level analyses explicitly examined the influence
of shared infrastructure.

Future iterations of the measurement framework could further improve
transparency by providing a publicly accessible project page or contact
mechanism through which organisations can identify the source of measurement
traffic and request exclusion from future measurement campaigns.

Despite these limitations, this study provides one of the first empirical
sector-level assessments of observable post-quantum cryptographic deployment
across UK organisations. By jointly analysing HTTPS and SMTP deployment,
infra\-str\-ucture-provider concentration, and certificate signature algorithms,
it establishes an empirical baseline against which future progress towards
post-quantum migration can be assessed and highlights the importance of
considering infrastructure-provider influence when interpreting Internet-scale
deployment measurements.
\section{Conclusion}
\label{sec:conclude}
This paper presented an empirical measurement study of observable
post-quantum cryptographic deployment across 4,665 UK organisations spanning
ten sectors. Using a sector-stratified dataset, we jointly analysed HTTPS and
SMTP STARTTLS infrastructures, attributed reachable endpoints to their
infrastructure providers, and examined leaf certificate signature algorithms.

Among reachable endpoints, 44.0\% of HTTPS services supported at least one
evaluated post-quantum key-exchange group, compared with 6.4\% of SMTP
services. Among organisations reachable over both protocols, observable PQC
support was approximately 16.9 times more likely over HTTPS than SMTP, and
only 144 organisations supported PQC across both services. Deployment was
overwhelmingly centred on the \texttt{X25519MLKEM768} hybrid group.

Although observable deployment differed significantly across organisational
sectors, the corresponding effect sizes were modest. Cross-validated modelling showed that
infra\-structure-provider identity was substantially more predictive than
organisational sector, achieving provider-only AUCs of 0.957 for HTTPS and
0.994 for SMTP. Deployment was also highly concentrated, with the largest
provider accounting for 69.7\% of PQC-supporting HTTPS endpoints and 98.4\%
of PQC-supporting SMTP endpoints. No post-quantum leaf-certificate signature
algorithms were observed.

These findings show that externally visible PQC support currently reflects
infrastructure-provider deployment decisions to a substantial extent and should
not be interpreted as a comprehensive measure of organisational migration
readiness. Furthermore, the absence of post-quantum certificate signatures
suggests that observable deployment remains focused on key establishment rather
than complete cryptographic migration. The measurements reported in this paper
provide an empirical baseline for future longitudinal studies of post-quantum
deployment across UK Internet-facing services.

\bibliographystyle{splncs04}
\bibliography{references}
\end{document}